\documentclass[twocolumn,showpacs,floatfix]{revtex4}
\usepackage{graphicx}
\begin{document}

\title {Exact diagonalization results for an anharmonically
trapped Bose-Einstein condensate}
\author{S. Bargi, G. M. Kavoulakis, and S. M. Reimann}
\affiliation{Mathematical Physics, Lund Institute of
Technology, P. O. Box 118, S-22100 Lund, Sweden}
\date{\today}

\begin{abstract}

We consider bosonic atoms that rotate in an anharmonic trapping
potential. Using numerical diagonalization of the Hamiltonian,
we identify the various phases of the gas as the rotational
frequency of the trap and the coupling between the atoms are
varied.

\end{abstract}

 \maketitle

\section{Introduction}

The behavior of confined, Bose-Einstein condensed atoms under
rotation has been studied extensively in recent years, both
experimentally
\cite{JILA,Madison,VortexLatticeBEC,HaljanCornell,JILAgiant},
as well as theoretically. One factor which is very crucial in
these studies is the form of the confinement. In most of the
experiments performed on cold gases of atoms, the trapping
potential is harmonic. It is interesting, however, that the
harmonic trapping is special in many respects, as we show
below. For this reason recent theoretical studies
\cite{Lundh,Fetter,tku,FG,KB,Emil,SPB,AA,EL,JK,JKL,KJB,Stringari,FZ,PKC,AK},
as well as the experiment of Ref.\,\cite{Dal}, have considered
traps with other functional forms, in which the trapping
potential grows more rapidly than quadratically at distances
far away from the center of the cloud. Such trapping potentials
introduce many novel phases.

Motivated by this observation, we examine the lowest state of a
Bose-Einstein condensate in a rotating anharmonic trap. Up to
now, all theoretical studies that have been performed on this
problem were based upon the mean-field approximation. However,
especially in the case of effective attractive interactions
between the atoms \cite{WGS,BM}, the use of the mean-field
approximation is questionable, as we explain in more detail
below. In this study, we present ``exact" results -- for small
numbers of atoms -- from numerical diagonalization of the
Hamiltonian, which goes beyond the mean-field approximation.

Experimentally it is possible to tune both the frequency of
rotation of the trap, as well as the strength of the
interatomic interaction. Motivated by these facts, we consider
here a fixed trapping potential and examine the various phases
and the corresponding phase diagram as function of the
rotational frequency and of the coupling.

In what follows we present our model in Sec.\,II. In Sec.\,III
we discuss the general structure of the phase diagram. In
Sec.\,IV we describe the details of the numerical
diagonalization. In Secs.\,V and VI we present and analyze our
results for effective repulsive and attractive interatomic
interactions, respectively. Finally, in Sec.\,VII we summarize
our main results.

\section{Model}

The Hamiltonian that we consider consists of the
single-particle part $h({\bf r}_i)$, and of the interaction
$V({\bf r}_i - {\bf r}_j)$, which is taken to be of the form of
a contact potential, $V({\bf r}_i - {\bf r}_j) = U_0 \delta
({\bf r}_i - {\bf r}_j)$,
\begin{equation}
     H = \sum_{i=1}^N h({\bf r}_i) + \frac 1 2 \sum_{i \neq j=1}^N
     V({\bf r}_i - {\bf r}_j).
\end{equation}
Here $N$ is the number of atoms, and $U_0 = 4 \pi \hbar^2 a/M$,
where $a$ is the scattering length for elastic atom-atom
collisions, and $M$ is the atom mass. The single-particle part
of the Hamiltonian has the usual form
\begin{equation}
     h({\bf r}) = - \frac {\hbar^2 } {2M}\nabla^2 +
     V_{\rm trap}({\bf r}),
\end{equation}
where the trap is assumed to be symmetric around the $z$ axis.
This is also taken to be the axis of rotation. In cylindrical
coordinates $\rho, \phi$, and $z$,
\begin{equation}
     V_{\rm trap}({\bf r}) = \frac 1 2 M \omega^2 \rho^2
     \left( 1 + \lambda \frac {\rho^2} {a_0^2} \right) + V(z).
\end{equation}
Here $\omega$ is the trap frequency of the harmonic part of
$V_{\rm trap}$, $\lambda$ is a dimensionless constant which is
the coefficient of the anharmonic term in $V_{\rm trap}$ (taken
to be much smaller than unity in this study, as is also the
case in the experiment of Ref.\,\cite{Dal}), and
$a_0=\sqrt{\hbar/M \omega}$ is the oscillator length. Along the
$z$ axis we assume that the trapping potential $V(z)$ is
sufficiently strong that the typical separation between
single-particle energy levels is much larger than the typical
interaction energy. This assumption implies that the motion
along the $z$ axis is frozen out and our problem is essentially
two-dimensional.

\section{General structure of the phase diagram}

As mentioned in the introduction, motivated by the experimental
situation, we here examine the phase diagram as function of the
rotational frequency $\Omega$ of the trap and of the strength
of the interaction. It is natural to measure $\Omega$ in units
of $\omega$ and the strength of the interaction $\sim n U_0$,
where $n$ is the atomic density, in units of the oscillator
energy $\hbar \omega$. The corresponding dimensionless ratio $n
U_0 / \hbar \omega$ is $\sim \sigma a$, where $\sigma = N/Z$ is
the density per unit length. Here $Z$ is the width of the
atomic cloud along the $z$ axis.

It is instructive to first consider the general structure of the
phase diagram \cite{KB,JKL,JK}. This consists of several
distinct phases. Let us start with repulsive interactions, $a >
0$. For a fixed $\Omega$, as $\sigma a$ increases, the total
energy eventually becomes dominated by the interaction energy, which
is minimized via the formation of singly quantized vortex
states. This is similar to vortex formation in liquid Helium, or in
harmonically trapped gases.

On the other hand, for a constant interaction strength, as the
rotational frequency increases -- and the single-particle part
is the dominant part of the Hamiltonian -- the total energy of
the gas is minimized via formation of multiply quantized
vortices. This is most easily seen if one thinks of the
effective potential felt by the atoms, which consists of
$V_{\rm trap}$ minus the centrifugal potential,
\begin{eqnarray}
     V_{\rm eff}({\bf r}) &=& V_{\rm trap}({\bf r}) -
     \frac 1 2 M \Omega^2 \rho^2 =
     \nonumber \\ &=& \frac 1 2 M (\omega^2-\Omega^2) \rho^2+
     \frac \lambda 2 M \omega^2 \frac {\rho^4} {a_0^2} + V(z).
\label{eff}
\end{eqnarray}
For the form of $V_{\rm trap}$ we have considered, $V_{\rm
eff}$ takes the form of a mexican-hat shape for $\Omega >
\omega$, and thus the atoms prefer to reside along the bottom
of this potential. This is precisely the density distribution
that corresponds to multiply-quantized vortex states. From
Eq.\,(\ref{eff}) it is also clear that for harmonic trapping,
$\lambda=0$, $\Omega$ cannot exceed $\omega$, since the atoms
will fly apart. For any other potential that grows more rapidly
than $\rho^2$ at large distances, the effective potential is
bounded, independently of $\Omega$. Finally, when both $\sigma
a$ and $\Omega/\omega$ are sufficiently large, there is a
third, mixed phase consisting of a multiply-quantized vortex
state at the center of the cloud, surrounded by
singly-quantized vortices. This configuration in a sense
compromises between the single-particle energy and the
interaction energy.

Turning to the case of effective attractive interactions,
$a<0$, for small enough $\sigma |a|$, the phase diagram is,
crudely speaking, symmetric with respect to $\sigma a$. The
lowest state of the gas is determined by the single-particle
part of the Hamiltonian, and thus the corresponding state still
consists of multiply quantized vortex states. For higher values
of $\sigma |a|$, these states are again unstable against the
formation of a combination of multiply quantized and singly
quantized vortices, as in the mixed phase described for
positive $\sigma a$. The only difference is that the
single-particle density distribution resembles that of a
localized ``blob'' rotating around the minimum of the effective
potential $V_{\rm eff}$ \cite{KJB}. For even higher values of
$\sigma |a|$, the phase that minimizes the energy is that of
the center of mass excitation first seen in harmonically
trapped atoms \cite{WGS,BM}. In this phase the angular momentum
is carried by the center of mass. The single-particle density
distribution resembles that of the ``mixed" phase (i.e., a
localized blob), although the structure of the many-body
wavefunction is very different. Finally, for sufficiently large
values of $\sigma |a|$ the gas collapses \cite{BP,UL,EM}.
Figure 1 shows a schematic phase diagram.
\begin{figure}[t]
\includegraphics[width=6.5cm,height=4.0cm]{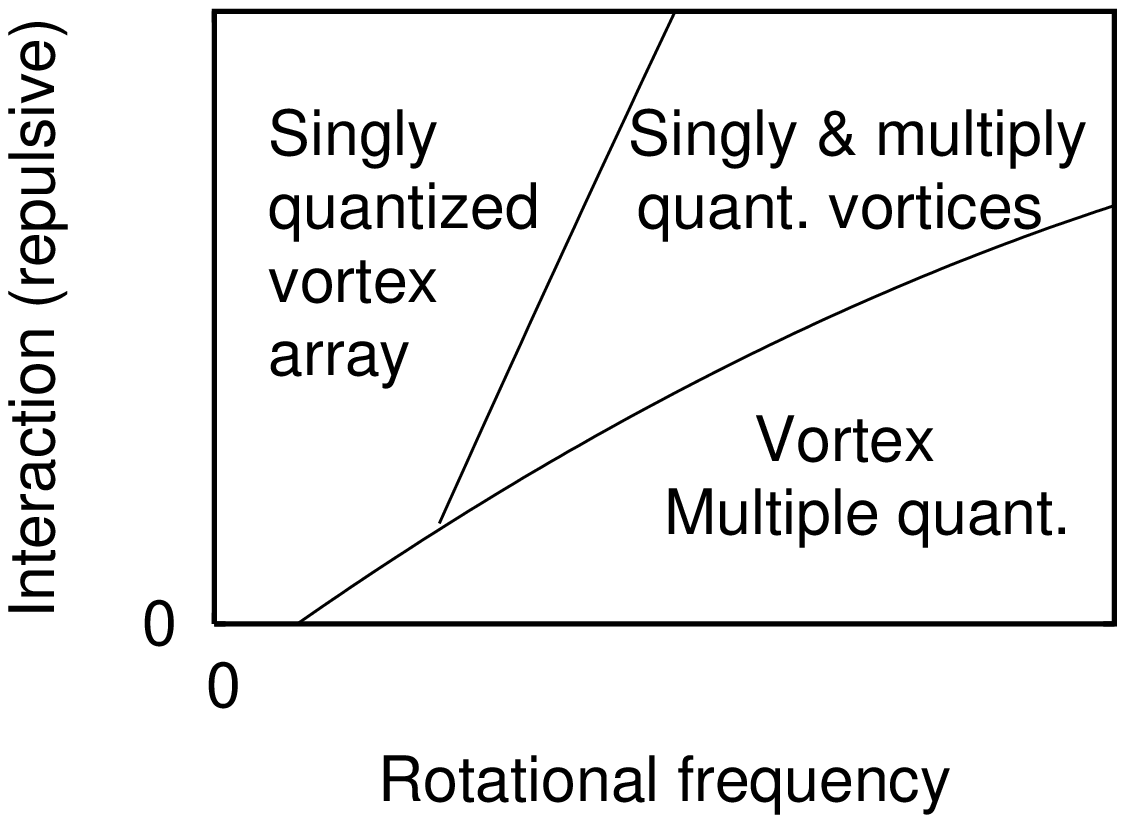}
\includegraphics[width=6.5cm,height=4.0cm]{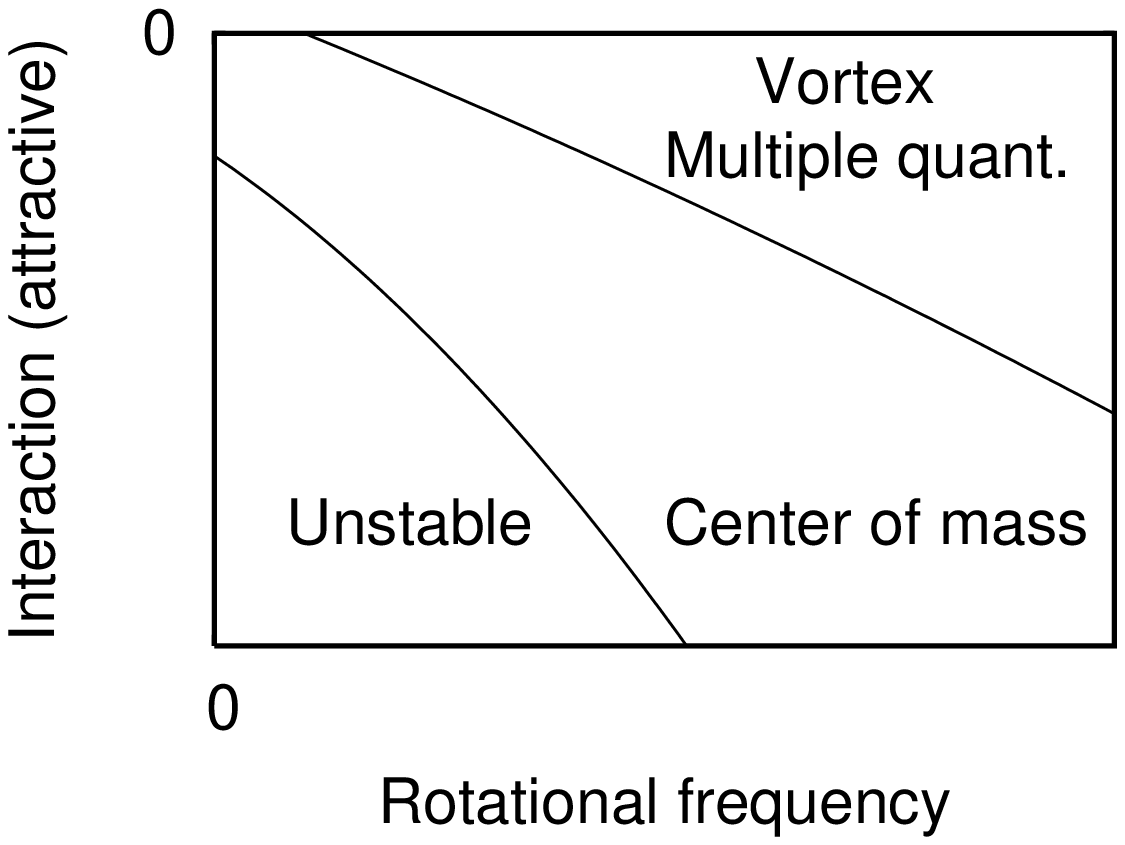}
\caption[]{Schematic phase diagram which shows the various
phases as the rotational frequency and the coupling are
varied.} \label{FIG1}
\end{figure}

\section{Numerical diagonalization of the Hamiltonian}

As mentioned earlier, in the present study we perform
numerical diagonalization of the Hamiltonian for a given
atom number $N$ and angular momentum $L \hbar$. To achieve
this, we first solve the single-particle
eigenvalue problem numerically,
\begin{equation}
    h \phi_{n_r,m} = E_{n_r,m} \phi_{n_r,m},
\end{equation}
using the Numerov method. The eigenfunctions and eigenenergies
are characterized by two quantum numbers, the number of radial
nodes $n_r$, and the number of quanta of angular momentum $m$.
Making the simplified assumption of weak interactions, $\sigma
|a| \ll 1$, we perform our calculation within the approximation
of the lowest Landau level, $n_r = 0$ and $m \geq 0$. States
with $n_r \neq 0$ or negative $m$ are higher in energy by a
term of order $\hbar \omega$. This assumption allows us to
consider relatively large values of $N$ and $L$. One should
remember that while for repulsive interactions $\sigma a$ can
have any value, for attractive interactions, $\sigma |a| \alt
1$, as otherwise the system collapses. Therefore, the lowest
Landau level approximation is more restrictive for repulsive
interactions, yet it gives a rather good description of the
various phases, even in this case.

Having found the single-particle eigenstates, we then set up
the Fock states, which are the eigenstates of the number
operator $\hat N$ and of the total angular momentum $\hat L$,
respectively. As a final step we set up the corresponding
Hamiltonian matrix between the various Fock states, which we
diagonalize numerically. The diagonal matrix elements consist
both of the single-particle part, as well as the interaction.
The only non-zero off-diagonal matrix elements result from the
interaction. The output of the Hamiltonian gives us not only
the lowest state, but the whole energy spectrum. Since our
calculation is performed for fixed angular momentum, in order
to work with a fixed $\Omega$ we consider the energy in the
rotating frame, which is given by the usual transformation
\begin{equation}
    H' = H - L \Omega.
\label{transf}
\end{equation}
Having found the many-body eigenstates and eigenvalues for some
range of $L$, then for a fixed $\Omega$ we identify the
eigenstate with the lowest eigenenergy according to
Eq.\,(\ref{transf}). The many-body eigenfunction that
corresponds to the lowest eigenenergy -- which is expressed in
terms of the Fock states -- is then analyzed in terms of the
occupancy of the single particle states. The calculation of the
occupancies is a trivial operation and, remarkably, each of the
phases described in the previous section is characterized by a
distinct distribution. This fact provides indisputable evidence
for the phases we expect to get from the arguments of the
previous section and from the predictions of mean-field.
\begin{figure}[t]
\includegraphics[width=6.5cm,height=4.0cm]{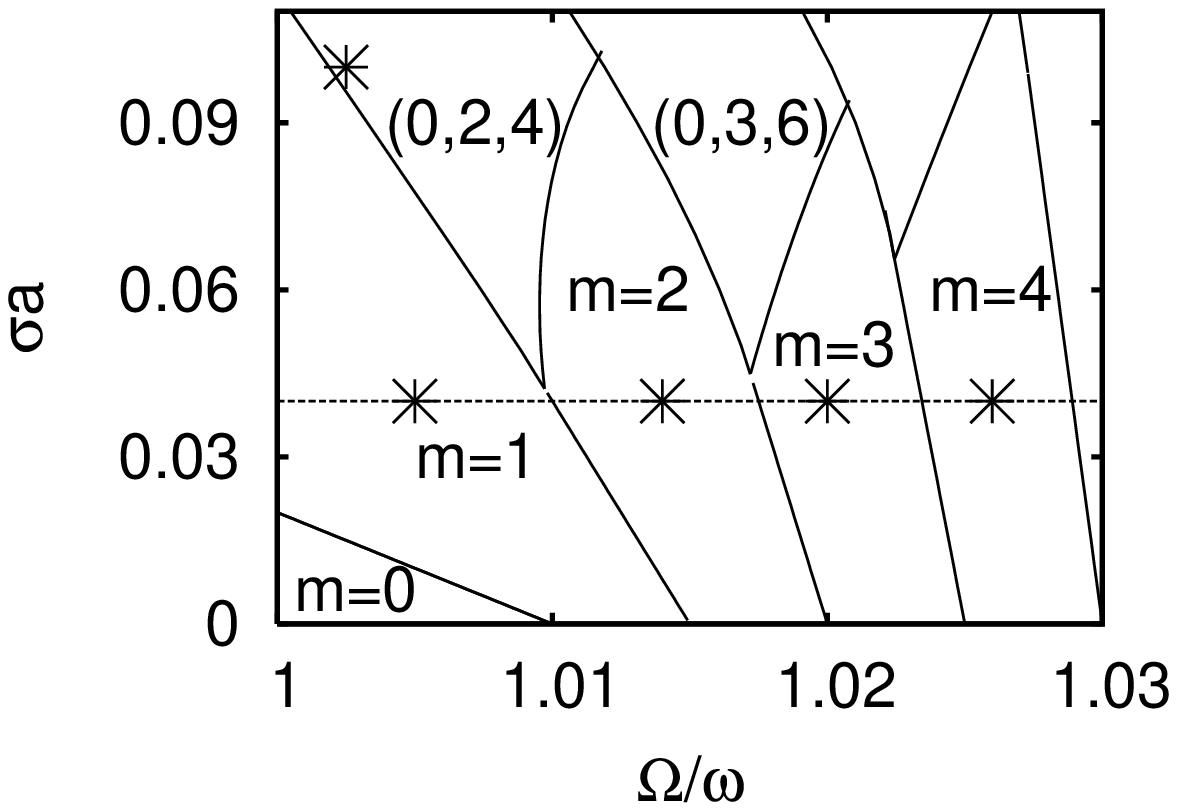}
\includegraphics[width=6.5cm,height=4.0cm]{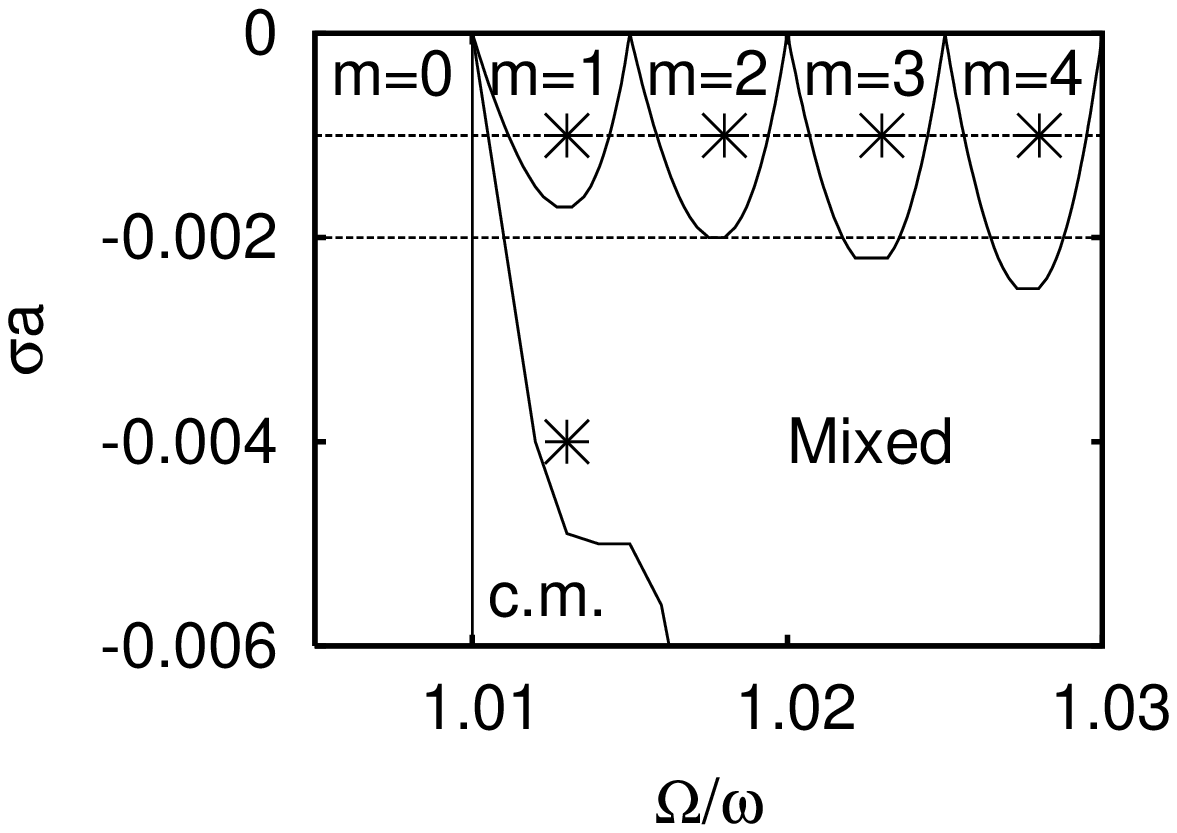}
\caption[]{The phase diagram calculated within the mean-field
approximation for $\lambda=0.005$ from
Refs.\,\cite{JKL,JK,KJB}. The ``c.m." phase is the center of
mass phase. The stars and the three dashed horizontal lines
represent some of the values of $\sigma a$ and $\Omega/\omega$
that have been examined and analyzed in the present study.}
\label{FIG2}
\end{figure}
\begin{figure}[t]
\includegraphics[width=5cm,height=4.0cm]{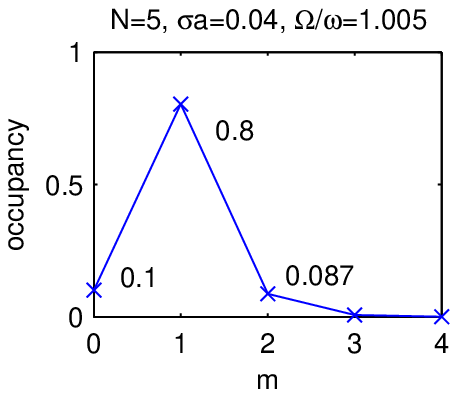}
\includegraphics[width=5cm,height=4.0cm]{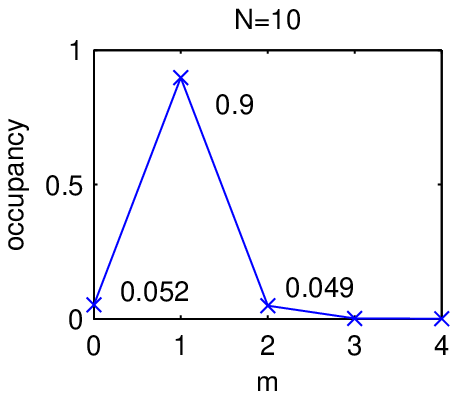}
\includegraphics[width=5cm,height=4.0cm]{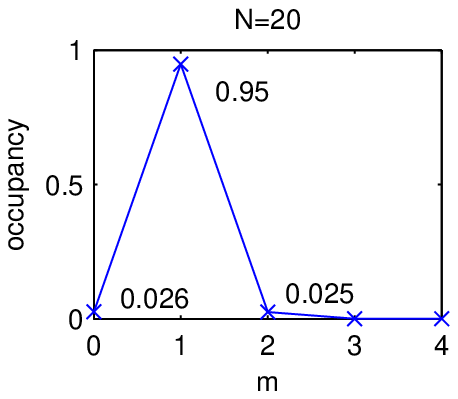}
\caption[]{The occupancy of the single-particle states of
angular momentum $m$ for $\lambda = 0.005$, $\sigma a= 0.04$,
$\Omega/\omega=1.005$, and for $N=5$ (higher), $N=10$ (middle),
and $N=20$ (lower). These graphs demonstrate that the occupancy
of all states with $m \neq 1$ is, to leading order, $1/N$. The
occupancy of the $m=1$ state is unity minus corrections of
order $1/N$ to leading order. In the mean-field approximation
this phase corresponds to a single vortex state.} \label{FIG3}
\end{figure}
\begin{figure}[h]
\includegraphics[width=9cm,height=10.0cm]{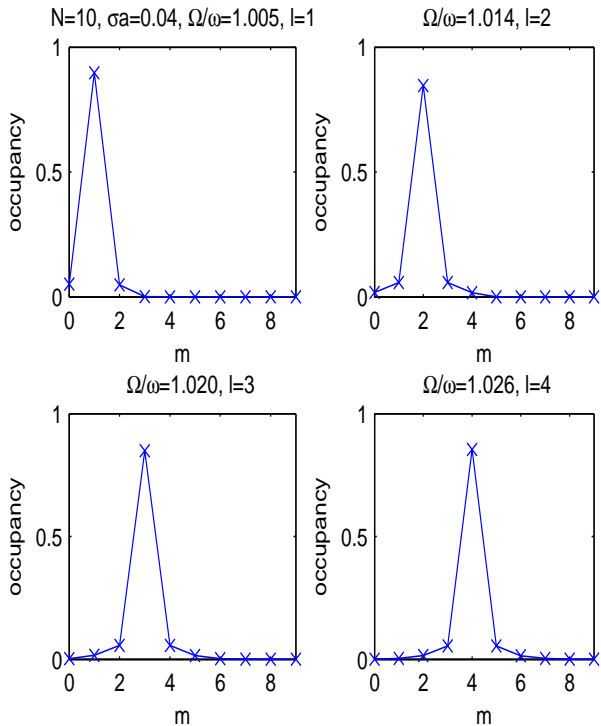}
\caption[]{The occupancy of the single-particle states of
angular momentum $m$ for $\lambda = 0.005$, $\sigma a= 0.04$,
$N=10$ atoms, and $\Omega/\omega=1.005$ (top left),
$\Omega/\omega=1.014$ (top right), $\Omega/\omega=1.020$
(bottom left), $\Omega/\omega=1.026$ (bottom right). The
angular momentum of the gas changes discontinuously between
successive values of $m$ as $\Omega$ increases. These phases
correspond to multiply-quantized vortex states.} \label{FIG4}
\end{figure}
\begin{figure}[h]
\includegraphics[width=6cm,height=5.0cm]{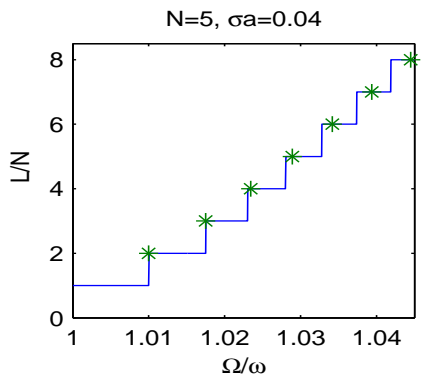}
\caption[]{The total angular momentum of the gas in its
lowest-energy state as a function of $\Omega/\omega$, for fixed
$\lambda = 0.005$, $\sigma a = 0.04$, and $N=5$. The stars
denote the result of a perturbative, mean-field calculation
from Ref.\,\cite{JK}.} \label{FIG5}
\end{figure}
\begin{figure}[h]
\includegraphics[width=6cm,height=5.0cm]{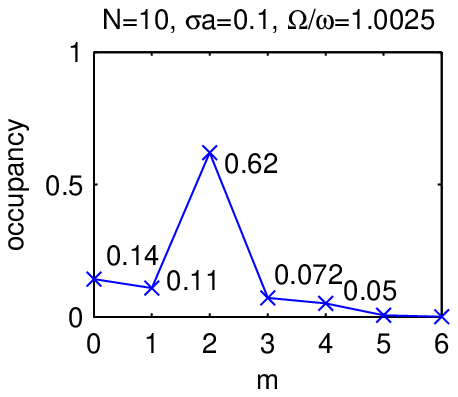}
\includegraphics[width=6cm,height=5.0cm]{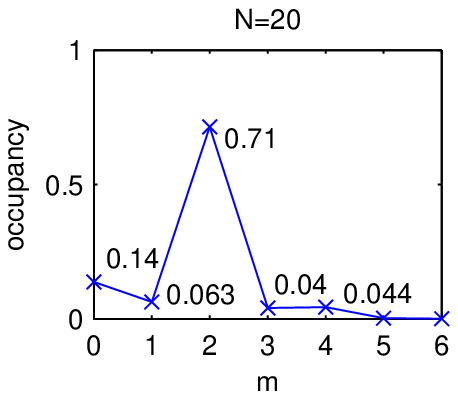}
\caption[]{The occupancy of the single-particle states of
angular momentum $m$ for $\lambda = 0.005$, $\sigma a = 0.1$,
$\Omega/\omega=1.0025$ and $N=10$ (higher), $N = 20$ (lower).
Here $L/N = 1.8$. These graphs demonstrate that the dominant
states are the ones with $m = 0, 2$, and 4, in agreement with
the mean-field calculations of Refs.\,\cite{JKL,JK}. In the
mean-field approximation this corresponds to a ``mixed" phase
discussed in the text.} \label{FIG6}
\end{figure}

\section{Results -- Repulsive interactions}

We turn now to the analysis of our results, starting with the
case of repulsive interactions. As we argued earlier, for
sufficiently weak interactions we expect the phase of multiple
quantization to have the lowest energy. For convenience we
compare our results with those of the phase diagram of
Ref.\,\cite{JKL}. We thus choose $\lambda = 0.005$ \cite{pd},
and evaluate the occupancy of the single particle states for
$\sigma a = 0.04$, $\Omega/\omega = 1.005$, for $N = 5$, 10,
and 20 atoms, as shown in Fig.\,3. This graph demonstrates
clearly that as $N$ increases, the occupancy of the $m=1$ state
approaches unity, while the occupancy of all other states tends
to zero (scaling as $1/N$ to leading order). In the mean-field
description the order parameter $\psi$ is simply
\begin{equation}
  \psi = \phi_{0,1},
\end{equation}
which represents a single vortex state located at the center of
the cloud.

In Fig.\,4 we calculate the occupancies of the single particle
states for $\lambda = 0.005$, $\sigma a = 0.04$ and for four
values of $\Omega/\omega=1.005, 1.014, 1.020$, and 1.026. This
graph confirms that indeed the system (in its lowest state)
undergoes discontinuous transitions between states of multiple
quantization of successive values of $m$. Figure 5 shows the
corresponding angular momentum per particle $l=L/N$ in the
lowest state of the gas versus $\Omega/\omega$, for a fixed
value of $\sigma a = 0.04$. This graph consists of
discontinuous jumps in $l$ as $\Omega$ increases. The asterisks
in this graph show the result of mean field \cite{JK}, which
compares well with our result for small $\Omega$. For larger
$\Omega$ the agreement becomes worse. This is because the
contribution of the anharmonic term to the energy was
calculated perturbatively in $\lambda$ in Ref.\,\cite{JK}. This
correction increases quadratically with $m$, and for larger
values of this quantum number, the perturbative result deviates
from the exact.

For repulsive interactions, as $\sigma a$ increases, mean-field
theory predicts that the system undergoes a continuous
transition from some multiply-quantized vortex state of angular
momentum $m_0$ to a ``mixed" state which, to leading order, is
a linear superposition of three states with angular momentum
$m_1$, $m_0$, and $m_2$, with $m_1+m_2 = 2 m_0$. Thus the order
parameter has the form
\begin{equation}
  \psi = c_{m_1} \phi_{0,m_1} + c_{m_0} \phi_{0,m_0}
  + c_{m_2} \phi_{0,m_2},
\end{equation}
where the coefficients are complex numbers. We have confirmed
that our method gives a similar transition, as shown in
Fig.\,6. In this figure we plot the occupancy of the
single-particle states for $\sigma a = 0.1$, $\Omega/\omega =
1.0025$, and $\lambda = 0.005$, for $N=10$ and $N=20$. From
these two graphs we see that the states with $m_1=0$, $m_0=2$,
$m_2=4$ are the dominant ones for large $N$. This is again in
agreement with the prediction of mean-field. Finally, the
angular momentum per particle $l=L/N$ is 1.8 in this state.

\begin{figure}[t]
\includegraphics[width=9cm,height=10.0cm]{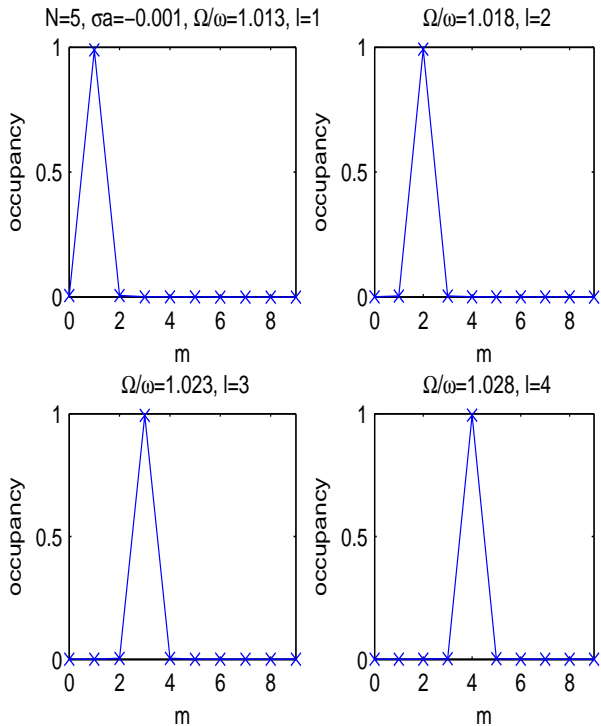}
\caption[]{The occupancy of the single-particle states of
angular momentum $m$ for $\lambda = 0.005$, $\sigma a =
-0.001$, $N=5$ atoms, and $\Omega/\omega=1.013$ (top left),
$\Omega/\omega=1.018$ (top right), $\Omega/\omega=1.023$
(bottom left), $\Omega/\omega=1.028$ (bottom right). In the
mean-field approximation this phase corresponds to
multiply-quantized vortex states.} \label{FIG7}
\end{figure}
\begin{figure}[t]
\includegraphics[width=6cm,height=5cm]{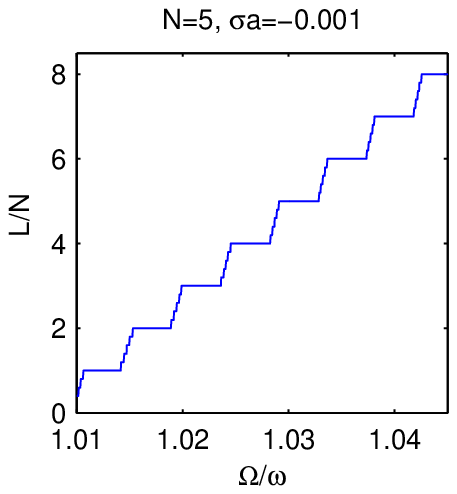}
\includegraphics[width=6cm,height=5cm]{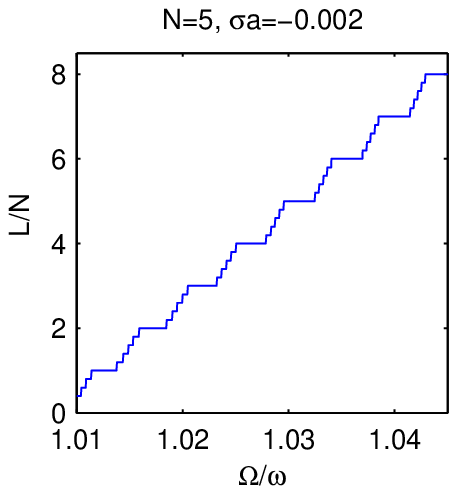}
\caption[]{The angular momentum per particle in the state of
lowest energy as a function of $\Omega/\omega$, for $\lambda =
0.005$, $N=5$, and $\sigma a = -0.001$ (higher), and $\sigma a
= -0.002$ (lower). The plateaus correspond to
multiply-quantized vortex states, which become more narrow as
$\sigma |a|$ increases (as also shown in Fig.\,2), in agreement
with the mean-field approximation \cite{KJB}. The small wiggles
in the curves between the plateaus are an artifact of the
discreetness of $L$ that we have considered.} \label{FIG8}
\end{figure}
\begin{figure}[t]
\includegraphics[width=5cm,height=4.0cm]{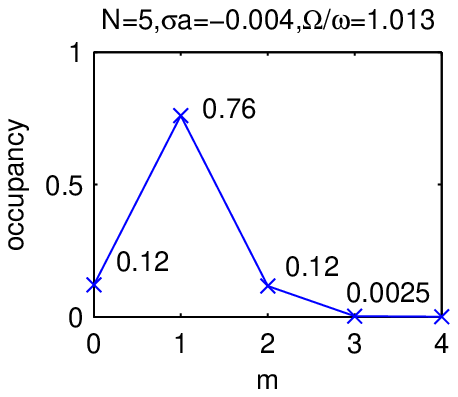}
\includegraphics[width=5cm,height=4.0cm]{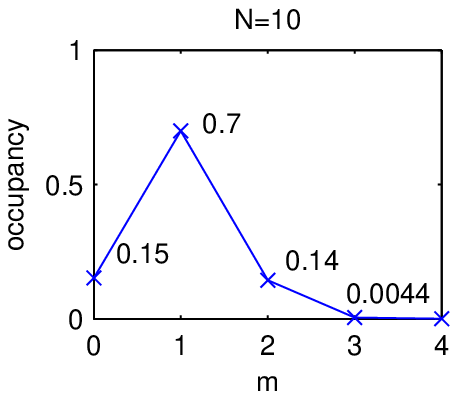}
\includegraphics[width=5cm,height=4.0cm]{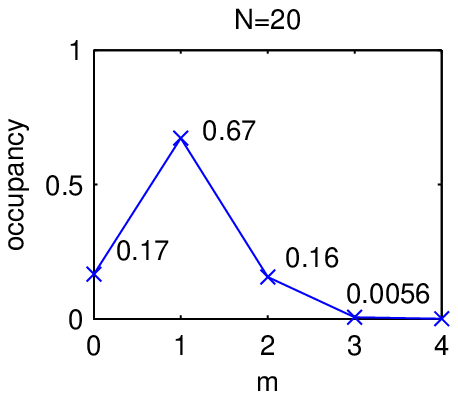}
\caption[]{The occupancy of the single-particle states of
angular momentum $m$ for $\lambda = 0.005$, $\sigma a =
-0.004$, $\Omega/\omega=1.013$, and for $N=5$ (higher), $N=10$
(middle), and $N=20$ (lower). Here $L/N = 1.0$. These graphs
demonstrate that the dominant states are the ones with $m = 0,
1$, and 2, with a small admixture of the state with $m=3$, in
agreement with Ref.\,\cite{KJB}. In the mean-field
approximation, this corresponds to the ``mixed" phase discussed
in the text.} \label{FIG9}
\end{figure}
\begin{figure}[t]
\includegraphics[width=9cm,height=10.0cm]{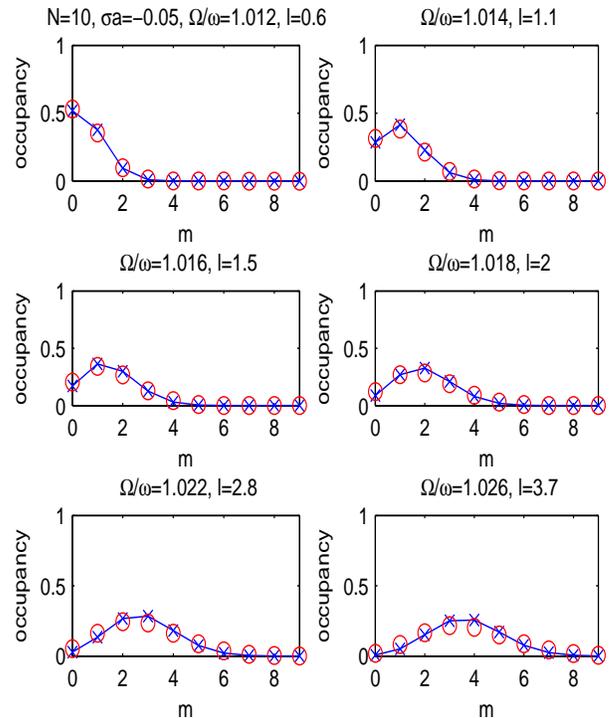}
\caption[]{The crosses show the occupancy of the
single-particle states of angular momentum $m$ for $\lambda =
0.005$, $\sigma a= -0.05$, $N=5$ atoms, and
$\Omega/\omega=1.012$ (top left), $\Omega/\omega=1.014$ (top
right), $\Omega/\omega=1.016$ (middle left),
$\Omega/\omega=1.018$ (middle right), $\Omega/\omega=1.022$
(bottom left), and $\Omega/\omega=1.026$ (bottom right). This
phase has a very large overlap with that of the center of mass
excitation. The circles show the results calculated from
Eq.\,(\ref{occ}).} \label{FIG10}
\end{figure}

\section{Results -- Attractive interactions}

For an effective attractive interaction between the atoms the
corresponding phase diagram is even richer. We have obtained
 evidence for all the phases that are expected, namely
the multiply-quantized vortex states, the mixed phase that
consists of singly and multiply quantized vortices, the one
involving center of mass excitation, and finally the unstable
one.

Figure 7 shows the occupancies of the single-particle states in
the lowest state of the gas for a fixed $\lambda = 0.005$ and
$\sigma a=-0.001$, and for four values of $\Omega/\omega =
1.013, 1.018, 1.023$, and 1.028. As in Fig.\,4, as $\Omega$
increases, we see that states of multiple quantization
$\phi_{0,m}$ with successive values of $m$ become
macroscopically occupied. In Fig.\,8 we plot the angular
momentum per particle $l=L/N$ versus $\Omega/\omega$ for
$\sigma a = -0.001$ and $-0.002$. Here there is a difference as
compared to repulsive interactions, which is also consistent
with the prediction of mean-field theory. In order for the gas
to get from some state of multiple quantization $m_0$ to
$m_0+1$ (corresponding to the plateaus in Fig.\,8), it has to
go through some ``mixed state", and thus these transitions are
continuous. Furthermore, the width (in $\Omega/\omega$) of the
plateaus decreases with increasing $\sigma |a|$. This effect is
clearly seen between the two graphs in Fig.\,8.

To get evidence for the mixed phase, we increase $\sigma |a|$,
$\sigma a = - 0.004$, and plot the occupancies of the single
particle states in Fig.\,9 for a fixed $\Omega/\omega=1.013$,
and $N=5, 10$, and 20. Here $L/N$ is 1.0. The dominant
single-particle states are the ones with $m = 0,1$, and 2, with
a very small admixture of $m=3$. Again, this is in agreement
with mean-field theory, which predicts that close to the phase
boundary, any multiply-quantized vortex state of strength $m_0$
is unstable against a state with three components of angular
momentum $m_0-1, m_0$, and $m_0+1$,
\begin{equation}
  \psi = c_{m_0-1} \phi_{0,m_0-1} + c_{m_0} \phi_{0,m_0}
  + c_{m_0+1} \phi_{0,m_0+1}.
\end{equation}
In such a linear superposition of single-particle states, the
corresponding single-particle density distribution may (and it
actually does) look very much like that of a localized blob
\cite{KJB}, although this is a phase involving vortex
excitation.

The phase of the center of mass excitation expected in this
problem \cite{KJB} was also clearly seen in our calculation for
even more negative values of $\sigma a$. This phase was first
discovered in harmonic traps \cite{WGS,BM}. Remarkably, the
many-body wavefunction can be written analytically and has a
very different structure as compared to the product form
assumed within the mean-field approximation. This fact makes
the validity of the mean-field approximation questionable for
attractive interactions.

The occupancy of the single-particle states can also be
expressed analytically. In Ref.\,\cite{WGS} it has been shown
that the occupancy $|c_m|^2$ of a single particle state with
angular momentum $m$ in a many-body state with center of mass
excitation of $L$ units of angular momentum and $N$ atoms is
\begin{equation}
    |c_m(L,N)|^2 = \frac {(N-1)^{L-m} L!} {N^L (L-m)! \, m!}.
\label{occ}
\end{equation}
The crosses in Fig.\,10 show the occupancy of the
single-particle states calculated numerically for $\sigma a =
-0.05$, $\lambda=0.005$, $N=10$, and six values of
$\Omega/\omega=1.012, 1.014, 1.016, 1.018, 1.022$, and 1.026.
The circles represent $|c_m|^2$ of Eq.\,(\ref{occ}). We
confirmed that the difference between the occupancies
calculated within our study and the prediction of
Eq.\,(\ref{occ}) decreases with increasing $N$ and increasing
$\sigma |a|$ (as long as the gas is stable). These data again
provide clear evidence for the phase of center of mass
excitation.

It should also be mentioned that the phase boundary between the
mixed phase and the one of center of mass excitation lies much
lower than the one calculated variationally in
Ref.\,\cite{KJB}. The reason is that the phase boundary of
Ref.\,\cite{KJB} (for these two specific phases) is just an
upper bound, and the present calculation is consistent with
this fact.

Finally, the unstable phase \cite{BP,UL,EM} also shows up in
our model. This is seen in the present calculation via the sign
of the chemical potential which becomes negative for a value of
$\sigma a \approx -1$. This result is inconsistent with the
assumption of weak interactions, $\sigma |a| \ll 1$, as the
approximation of lowest Landau level breaks down. To get a
quantitatively accurate description of this instability one
would have to include higher Landau levels.

\section{Summary}

To summarize, in the present study we considered bosonic atoms
that rotate in an anharmonic trap. We attacked this problem
with the method of numerical diagonalization of the
Hamiltonian, which goes beyond the mean-field approximation but
also forced us to consider small numbers of atoms. Apart from
the assumption of weak interatomic interactions, our results
are exact. We got evidence for all the phases that are
expected, namely vortex excitation of multiple quantization,
single quantization, and mixed; center of mass motion, and
finally the unstable phase. The location of the phase
boundaries were consistent with those calculated within the
mean field approximation.

\acknowledgements We thank Yongle Yu for his assistance on the
numerical calculations. This work was partly financed by the
European Community project ULTRA-1D (NMP4-CT-2003-505457), the
Swedish Research Council, and the Swedish Foundation for
Strategic Research.


\begin{thebibliography} {99}
\bibitem{pd} In Ref.\,\cite{JKL} the value of $\lambda$ that was
used was 0.05, however here we use $\lambda=0.005$. To compare the
present results with those of this reference, the axes of the phase
diagram of Ref.\,\cite{JKL} have been rescaled. This rescaling is
allowed because of the universality of the phase diagram in the
limit of small $\lambda$, as shown in this reference.
\end{thebibliography}
\end{document}